\documentclass[preprint]{aastex}

\begin{document}
\title{Recent X-ray Observations of Disk Galaxies: Tracing the Dynamic Interstellar Medium}
 \author{Q. Daniel Wang}
\affil{University of Massachusetts, Amherst}

\begin{abstract}

I review recent results from our deep ROSAT and Chandra observations of two 
galaxies, M101 and NGC 4631, in fields of exceptionally low Galactic 
extinction. Large amounts of X-ray-emitting gas are detected in these galaxies.
Such gas is produced primarily in massive star forming regions and
have an average characteristic temperature of a few times $10^6$~K. Cooler
gas ($\sim 10^6$~K) is found typically outside galactic disks and may 
represent outflows from blown-out superbubbles. Propagation of star formation,
driven by the expansion of hot gas, appears to be operating in giant HII
complexes. A substantial fraction of photo-evaporated gas in such complexes 
may be mass-loaded into hot gas, which explains their large X-ray 
luminosities. These processes likely play an important role in determining the
global properties of the interstellar medium, especially the disk/halo 
interaction.
\end{abstract}

\section{Introduction}

X-ray observations of nearby disk galaxies are important to the understanding 
of the hot interstellar medium (ISM) and its role in galaxy evolution. We 
are conducting an extensive study of the face-on spiral galaxy M101 based 
partly on 
an ultra-deep (230 ks) ROSAT HRI image,  and we have also obtained a Chandra 
observation on the edge-on galaxy NGC4631. These observations, together with 
data in other wavelength bands, provide us with complementary 
perspectives of hot gas and its interaction with other galactic 
components in late-type disk galaxies. 

\section{M101}

This face-on galaxy allows us to study the production of 
hot gas not only under normal circumstances but also under localized starburst
conditions. Diffuse $\sim 0.25$~keV emission is present in the central 
($R \la 10$~kpc) region of the galaxy (Snowden \& Pietsch 1995).  Enhanced 
X-ray emission is also detected in several giant HII complexes (GHCs; Williams 
\& Chu 1995).  Wang et al. (1999) have reported the results on discrete 
X-ray sources detected in the M101 field. But some of these sources
apparently represent peaks of extended X-ray-emitting features.

\begin{figure}
\plotone{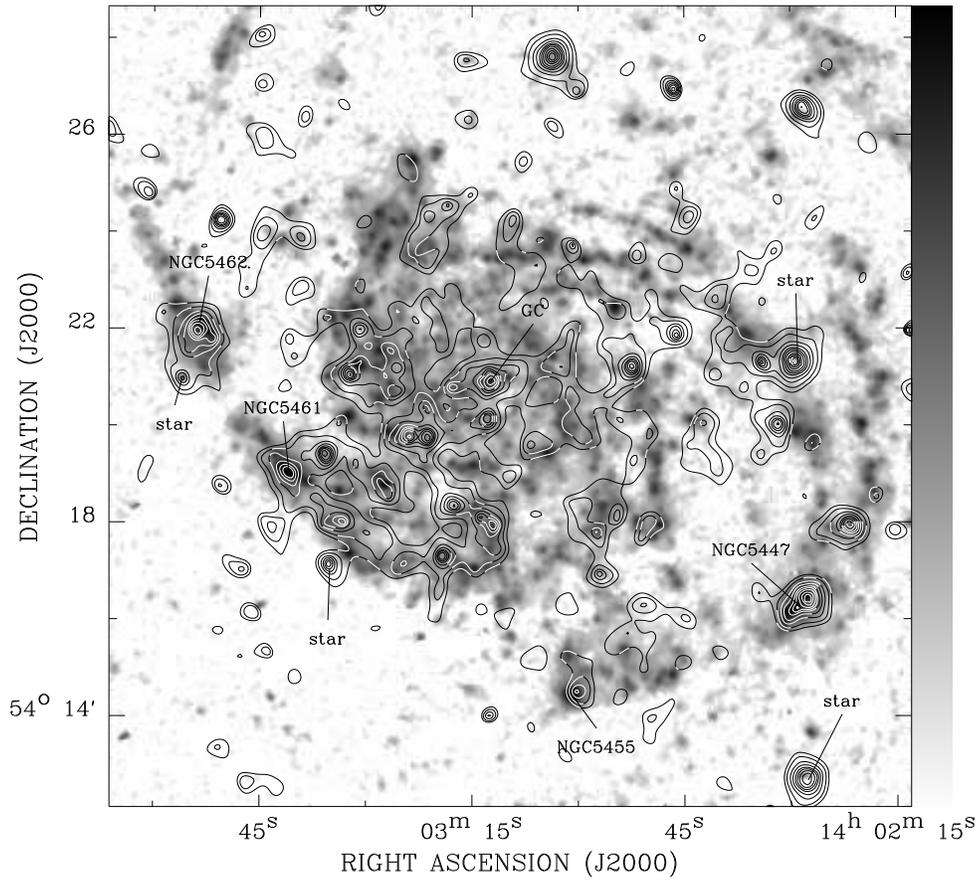}
\caption{ RHRI contours (Wang et al. 1999) overlaid on an 
H$\alpha$ image of M101.}
\end{figure}

Hot gas (a few times $10^6$~K) as traced by the RHRI correlates well with 
recent star forming regions in the galaxy (e.g., Fig. 1).
In particular, enhanced diffuse X-ray emission features line up well
with the active southeast arm, which harbors such GHCs as NGC 5461
and NGC 5462, which are more luminous than the 30 Dor complex in the LMC. 
We further find that the radial distribution of 
``diffuse'' X-ray emission is nearly identical to that of the far-UV 
radiation and is substantially flatter than that of optical light, which 
is produced chiefly by old stellar populations. Thus the hot gas originates 
primarily in recent star formation regions. In comparison,
the cooler gas ($\sim 10^6$~K) as traced by the ROSAT PSPC 0.25 keV image 
(Snowden \& Pietsch 1995) is seen only in the inner 4$^\prime$ radius,
consistent with an origin in the galaxy's corona or a combination of blown-out
bubbles.

\begin{figure}
\plotone{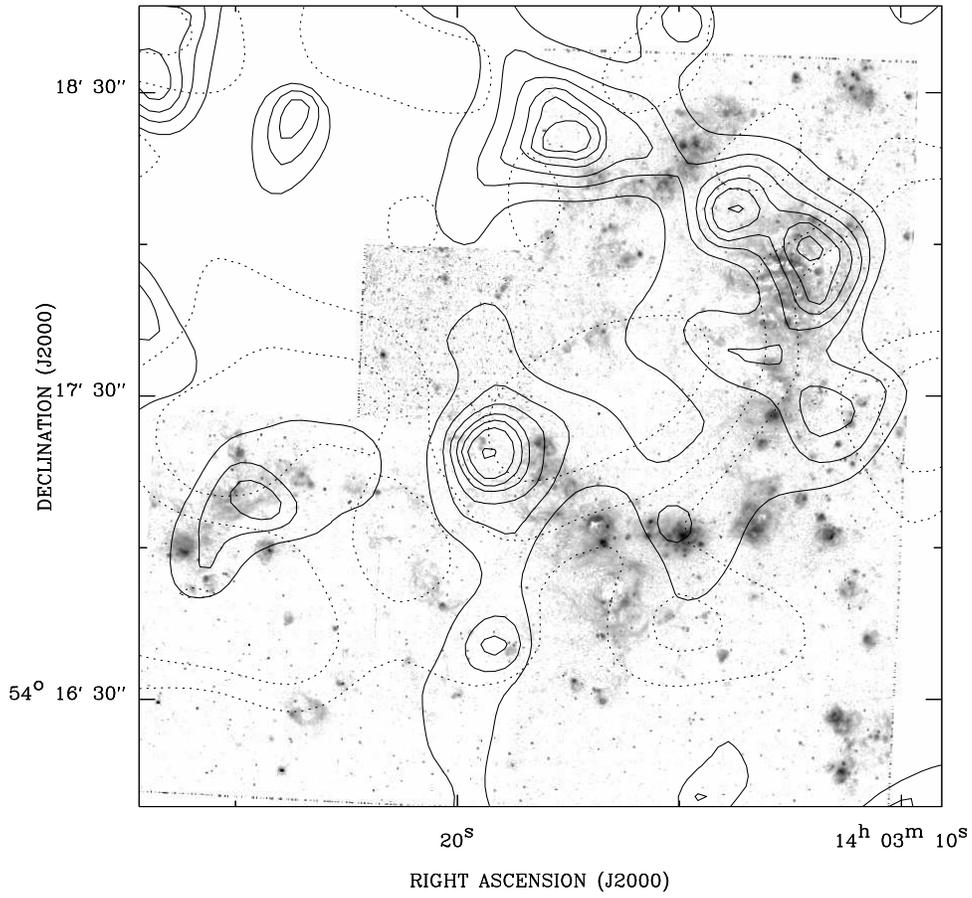}
\caption{A close-up of a region 
imaged by HST WFPC2 (H$\alpha$). The solid contours are the RHRI contours, 
while the dotted ones represent the PSPC 0.25 keV intensity.}
\end{figure}

Fig. 2 provides a close-up of a region at the western end of the two 
H$\alpha$ branches of the southeast spiral arm (Fig. 1). This region happens to
be covered by archival HST WFPC2 observations. An apparent ring of 
HII regions coincides spatially with an HI supershell, as proposed by Kamphuis (1993). Whereas the RHRI emission is
primarily associated with the HII ring, the softer PSPC 0.25~keV
emission peaks in the interior of the ring. Thus, the X-ray-emitting
gas within the ring appears to be cooler than that in the HII
regions. If the HI supershell is indeed powered by the mechanical
energy input from massive stars, its dynamic age is several times $10^7$ yrs,
about an order of magnitude greater than that of the HII regions. This 
can naturally be explained by the propagation of star formation,
which is likely triggered by the expansion of the supershell. 

The HST image also presents morphological evidence for
outflows from HII regions into the interior of ring.  These
frothy HII regions appear being torn apart, being stripped
away, and mass-loading the interior of the supershell.
Similar feedback activities are also 
evident in the GHC NGC~5462.

\section{NGC4631}

\begin{figure}
\plotone{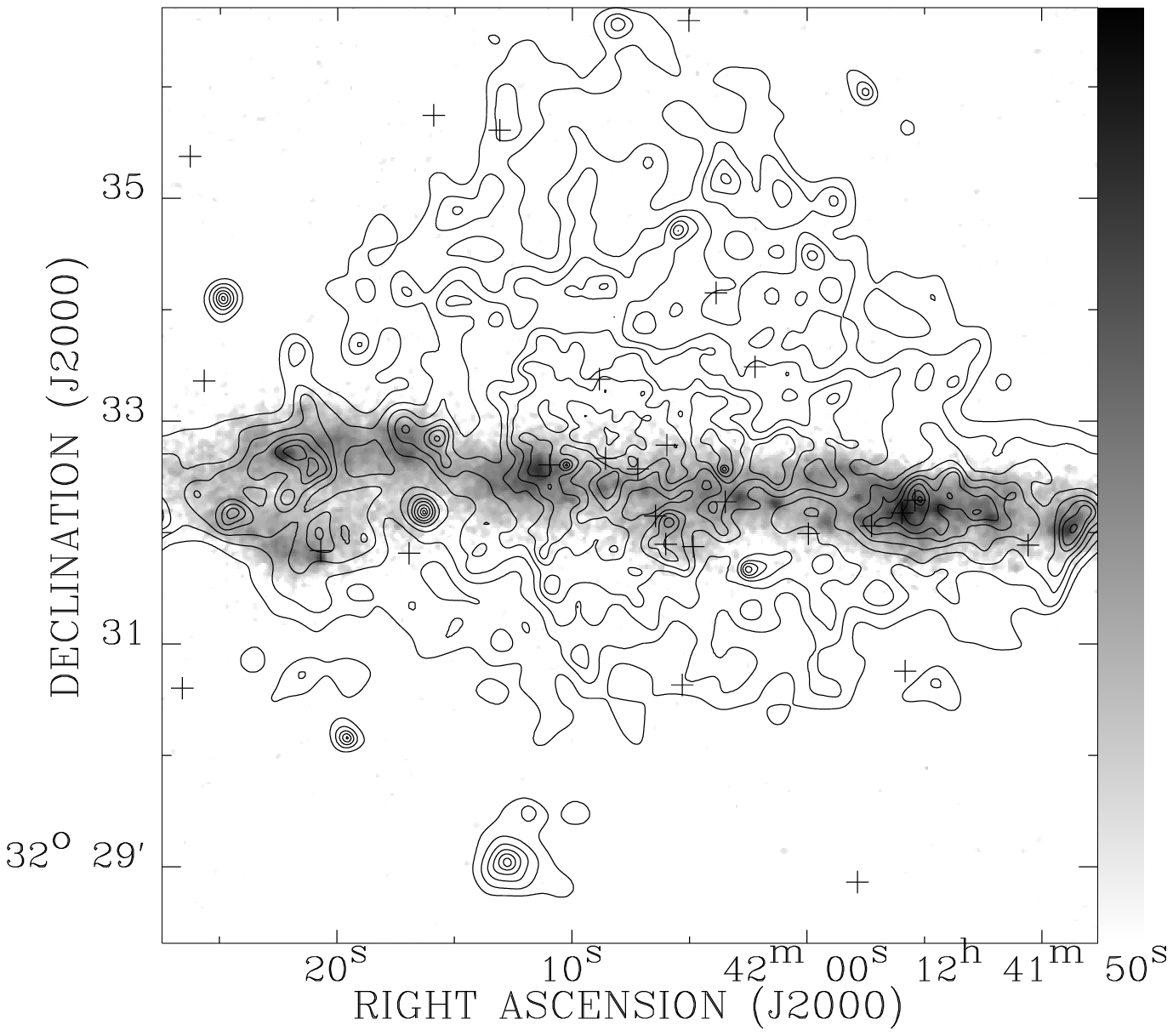}
\caption{Diffuse X-ray intensity contours overlaid on a far-UV image of NGC4631. The X-ray
data are from the Chandra ACIS-S in the 0.5-1.5 keV band. Point-like sources as
marked by pluses have been excised.}
\end{figure}

	ROSAT observations have shown that X-ray emission, particularly
in the 0.25 keV band, extends beyond both the cold and warm gas disks
of the galaxy (Wang et al. 1995). The Chandra ACIS-S observation
of the galaxy provides us with a new opportunity to further the 
study. While the analysis of the data is still on-going, Fig. 3 shows
unambiguously the presence of extraplanar diffuse X-ray emission. The
X-ray morphology resembles the radio halo of the galaxy (e.g., 
Hummel et al. 1991), indicating
a close connection among outflows of hot gas, cosmic rays, and magnetic field.
A spatially-resolved spectroscopy will further enable us to study 
the detailed chemical and thermal properties of the hot gas, constraining 
the dynamics of the corona. The results will have strong implications for 
our understanding of the feedback from galaxies as well as their 
structure and evolution.


\begin{references} 
\reference{} Hummel, E., Beck, R., \& Dahlem, M. 19991, A\&A, 248, 23
\reference{} Kamphuis, J. J. 1993, Ph.D. thesis 
\reference{} Snowden, S. L., \& Pietsch, W. 1995, ApJ, 452, 627 
\reference{} Wang, Q. D., et al. 1995, ApJ, 439, 381
\reference{} Wang, Q. D., Immler, S., \& Pietsch, W. 1999, ApJ, 523, 121
\reference{} Williams, R., \& Chu, Y.-H. 1995, ApJ, 439, 132 
\end{references}
\end{document}